\documentclass[12pt]{article}
\usepackage{graphicx,amsmath}
\usepackage{units}

\newlength{\dinwidth}
\newlength{\dinmargin}
\def\lapproxeq{\lower .7ex\hbox{$\;\stackrel{\textstyle                                                    
<}{\sim}\;$}}                                                    
\def\gapproxeq{\lower .7ex\hbox{$\;\stackrel{\textstyle                                                    
>}{\sim}\;$}}    
\def\k{\vec{k}_t} 
\def\kk{\vec{k}'_t} 
\def\b{\vec{b}}
\def\bb{\vec{b}'}     
\def\be{\begin{equation}}                                                    
\def\ee{\end{equation}}

\begin{document}

\begin{flushright}                                                    
IPPP/12/39  \\
DCPT/12/78 \\                                                    
\today \\                                                    
\end{flushright} 

\vspace*{0.5cm}

\begin{center}
{\Large \bf Diffractive Physics}

\vspace*{1cm}
                                                   
A.D. Martin$^a$\footnote{Plenary talk at 6th Int. Conf. on Quarks and Nuclear Physics, Ecole Poly., Palaiseau, Paris, April 2012}, H. Hoeth$^a$, V.A. Khoze$^{a,b}$, F. Krauss$^a$, M.G. Ryskin$^{a,b}$ and K. Zapp$^a$\\                                                    
                                                   
\vspace*{0.5cm}                                                    
$^a$ Institute for Particle Physics Phenomenology, University of Durham, Durham, DH1 3LE \\                                                   

$^b$ Petersburg Nuclear Physics Institute, NRC Kurchatov Institute, Gatchina, St.~Petersburg, 188300, Russia \\          
                                                    
\vspace*{1cm}                                                    
                                                    
\begin{abstract}                                                    
`Soft' high-energy interactions are clearly important in $pp$ collisions. Indeed, these events are dominant by many orders of magnitude, and about 40$\%$ are of diffractive origin; that is, due to elastic scattering or proton dissociation. Moreover, soft interactions unavoidably give an underlying component to the rare `hard' events, from which we hope to extract new physics. Here, we discuss how to quantify this contamination. First we present a brief introduction to diffraction. We emphasize the different treatment required for proton dissociation into low- and high-mass systems; the former requiring a multichannel eikonal approach, and the latter the computation of triple-Pomeron diagrams with multi-Pomeron corrections. Then we give an overview of the Pomeron, and explain how the QCD (BFKL-type) Pomeron is the natural object to continue from the `hard' to the `soft' domain. In this way we can obtain a partonic description of soft interactions. We introduce the so-called KMR model, based on this partonic approach, which includes absorptive multi-Pomeron corrections that become increasingly important as we proceed further into the soft domain. This model is able to describe total, elastic and proton dissociation data, and to predict the survival probability of large rapidity gaps to soft rescattering --- in terms of a few physically-motivated parameters. However, more differential phenomena, such as single particle $p_t$ distributions, can only be satisfactorily described if hadronization effects are included. This is achieved by incorporating the KMR analytic approach into the SHERPA Monte Carlo framework. It allows a description of soft physics and diffraction, together with jet physics, in a coherent, self-consistent way. We outline the structure, and show a few results, of this Monte Carlo, which we call SHRiMPS, for reasons which will become clear.

\end{abstract}                                                        
\vspace*{0.5cm}                                                    
                                                    
\end{center}

\section{Introduction}

There is no unique definition of diffraction \cite{review}.  We may say diffraction is elastic (or quasi-elastic) scattering caused, via {\it $s$-channel} unitarity, by the absorption of components of the wave functions of the incoming particles. For example at the LHC, there are processes $pp \to pp,~pX,~XX$, where neither, one, or both, protons are allowed to dissociate into a system $X$ with the quantum numbers of the proton. This definition is only a useful definition for quasi-elastic processes, but not for dissociation into high-mass systems.

An alternative definition is that a diffractive process is characterized by a large rapidity gap (LRG), which is caused by {\it $t$-channel} `Pomeron' exchange. However this definition is only good for very LRG events; otherwise the gap can be due to secondary Reggeon exchange or by fluctuations in the hadronization process.

Let us return to $s$-channel unitarity: $SS^\dagger =I$ with $S\equiv I+iT$, so that $T-T^\dagger = iT^\dagger T$. For fixed impact parameter, $b$, if we sandwich the relation between elastic states we obtain
\begin{equation}
2{\rm Im}T_{\rm el}(s,b)~=~|T_{\rm el}(s,b)|^2+G_{\rm inel}(s,b)~~~~~{\rm which~is~satisfied~by}~~~~~T_{\rm el}(s,b)=i(1-e^{-\Omega /2}). 
\label{eq:unitarity}
\end{equation}
At high energies the elastic amplitude is dominantly imaginary, so the eikonal/opacity, $\Omega$ is real, and greater than or equal to zero.
The total, elastic and inelastic cross sections are easily expressed in terms of $\Omega$. For example
\begin{equation}
\frac{d^2\sigma_{\rm inel}}{d^2b}~=~\frac{d^2\sigma_{\rm tot}}{d^2b}-\frac{d^2\sigma_{\rm el}}{d^2b}~=~2{\rm Im}T_{\rm el}-|T_{\rm el}|^2=1-e^{-\Omega},
\end{equation} 
so $e^{-\Omega (s,b)}$ is the probability of no inelastic interaction at $b$. The textbook example of how absorption (that is the $G_{\rm inel}$ term in (\ref{eq:unitarity})) gives rise to elastic scattering, is scattering on a black disc with Im$T_{\rm el}=1$ for $b<R$; then we have  $\sigma_{\rm inel}=\pi R^2$, which induces 
$\sigma_{\rm el}=\pi R^2$, so $\sigma_{\rm tot}=2\pi R^2$.

\begin{figure}[h]
\begin{center}
\resizebox{1.5\textwidth}{!}{
\includegraphics{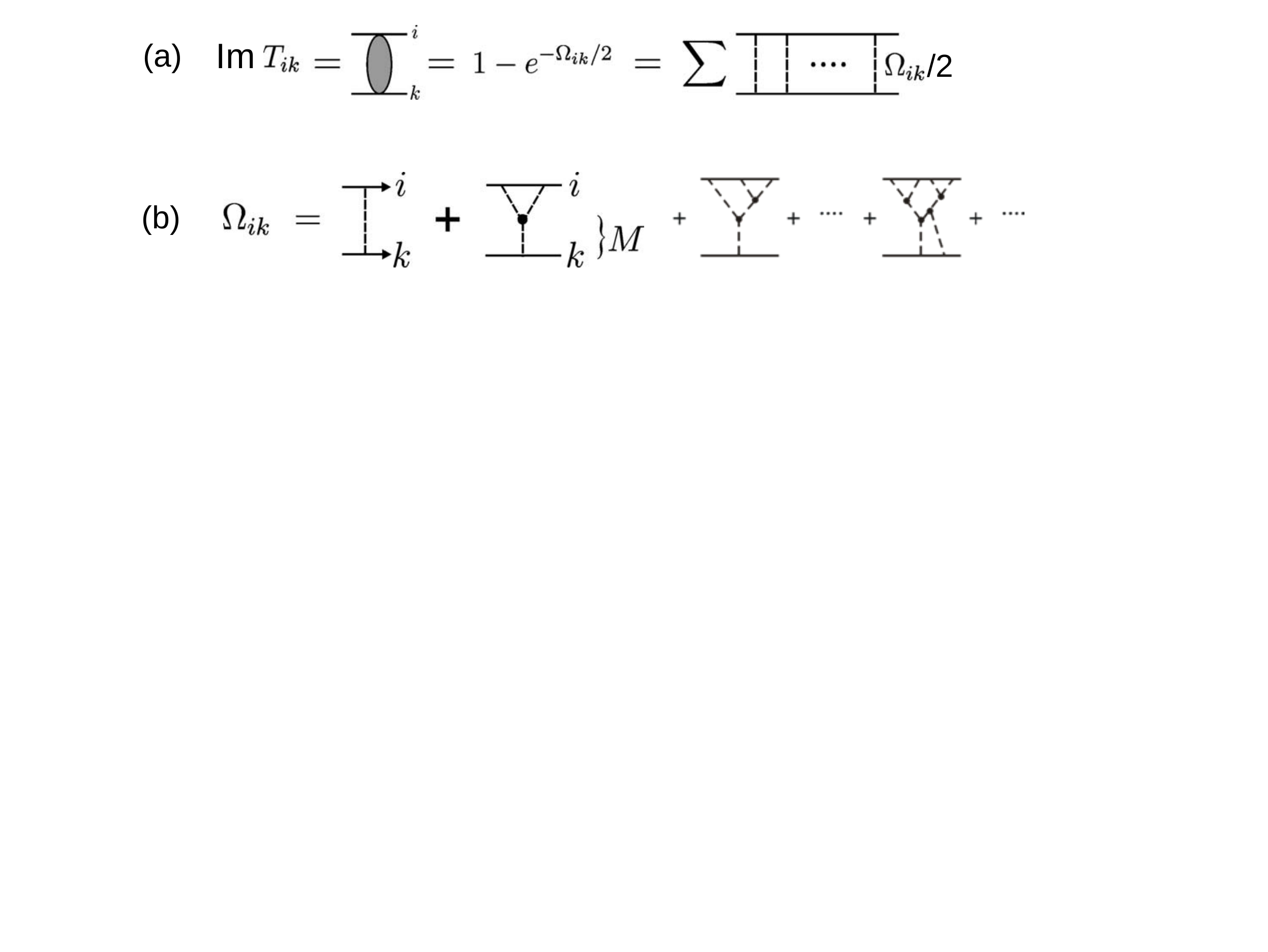}\hspace{12cm}
}
\vspace{-9cm}
\caption{\sf High energy $pp$ scattering, showing (a) the multichannel eikonal formalism to allow for low-mass proton dissociation, and (b) the triple-Pomeron diagram and multi-Pomeron corrections which account for high-mass dissociation. }
\label{fig:1}
\end{center}
\end{figure}

So much for elastic unitarity. What about proton dissociation? To include dissociation into low-mass states we have the Good-Walker formalism. We introduce combinations of $p,p^*,..$ (the so-called diffractive eigenstates $i,k,..$) which only undergo `elastic' scattering. Thus (\ref{eq:unitarity}) is generalized into a multichannel eikonal formalism, sketched in Fig. \ref{fig:1}(a).

What about dissociation into systems of high-mass $M$? High $M$ production may be represented by Pomeron exchange, giving rise to the triple-Pomeron exchange diagram, shown symbolically in Fig. \ref{fig:1}(b) with its multi-Pomeron absorptive corrections.
\vspace{-0.3cm}

\section{Why study diffraction?}

Well, first, there is its intrinsic interest; the LHC offers the possibility to probe the asymptotic behaviour of the $pp$ interaction. Here, we are concerned with a more practical reason. About 40$\%$ of the events at the LHC are due to diffraction --- elastic scattering and proton dissociation. Together with other `soft' interactions, these events dominate by many orders of magnitude. Moreover the LHC detectors do not have 4$\pi$ geometry. Although experimental triggers are able to select the rare `hard' interaction events at the LHC, from which we seek evidence of `new physics', they contain particles from the `soft' underlying interaction. It is therefore of great importance to construct a Monte Carlo, which includes diffraction, and which merges soft and hard interactions in a coherent self-consistent way.

 A simple example, illustrating the need of such a Monte Carlo, is related to the measurement of the energy scale 
of jets - additional hadronic activity beyond the standard parton shower plus
hadronization approach clearly may change measurements of, say, the $p_\perp$
distribution of jets.
Another example is the potentially informative exclusive production processes at the LHC where one can study the quasi-elastic hard subprocess in a very clean experimental environment. Unfortunately, in the present LHC experiments we are unable to determine the full kinematics of these exclusive events.  Such events are selected simply by the existence of a Large Rapidity Gap (LRG). Thus we need a reliable MC generator to estimate the mixture of the pure exclusive process with the processes where an incoming proton dissociates into either a low-mass or a high-mass system.  Such a Monte Carlo would also be valuable for interpreting high energy cosmic ray data as obtained, for example, in the Auger experiment.

To construct the sought-after Monte Carlo, which describes soft and hard interactions in a unified framework, we first need a partonic model of the Pomeron.

\section{The Pomeron}

Conventionally, `soft' and `hard' high-energy $pp$ interactions are described in terms of  different formalisms.  High-energy soft interactions are described by Reggeon Field Theory \cite{RFT} with a phenomenological (soft) Pomeron, whereas for hard interactions we use a QCD partonic approach. In perturbative QCD (pQCD), the Pomeron is associated with the BFKL vacuum singularity \cite{book}. However, the two approaches appear to merge naturally into one another.  That is, the partonic approach seems to extend smoothly into the soft domain. 

The BFKL equation describes the development of the gluon shower as the momentum fraction, $x$, of the proton carried by the gluon decreases.  That is, the evolution parameter is ln$(1/x)$, rather than the ln$k_t^2$ evolution of the DGLAP equation. 
Formally, to justify the use of pQCD, the BFKL equation should be written for gluons with sufficiently large $k_t$. However, it turns out that, after accounting for next-to-leading ln$(1/x)$ corrections and performing an all-order resummation of the main higher-order contributions \cite{bfklresum}, the intercept of the BFKL Pomeron depends only weakly on the scale for reasonably small scales. The intercept is found to be $\alpha_P(0)= \sim 1.3$ over a large interval of smallish $k_t$.
Thus the BFKL Pomeron is a natural object to continue from the `hard' domain into the `soft' region.

\begin{figure}[htb]
\begin{center}
\vspace{-2.5cm}
\resizebox{0.8\textwidth}{!}{
\includegraphics{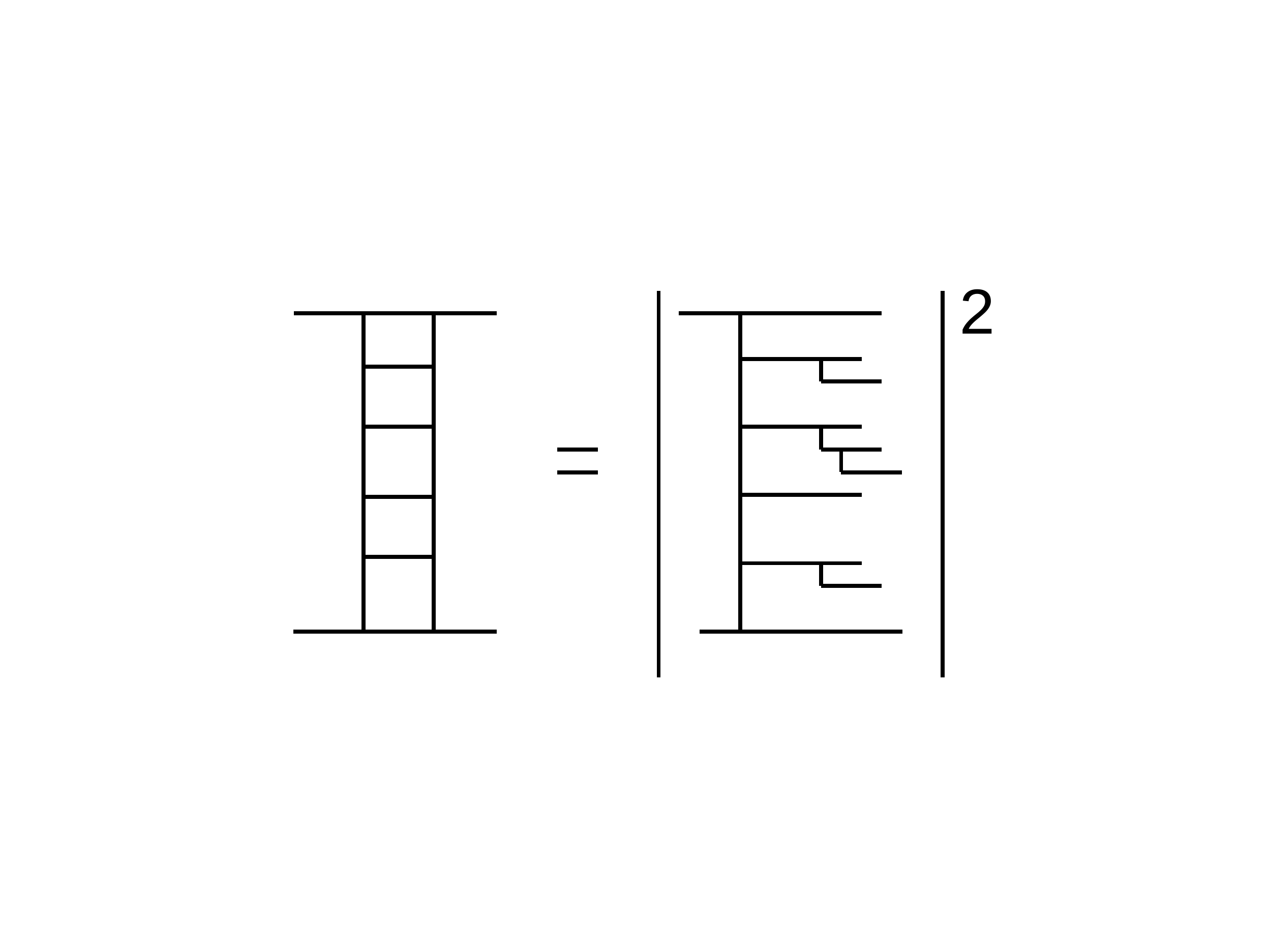}\hspace{1cm}
}
\vspace{-3cm}
\caption{\sf The cascade structure of a gluon ladder. The BFKL or QCD Pomeron is the sum of ladder diagrams, each with a different number of rungs.}
\label{fig:cascade}
\end{center}
\end{figure}

In terms of Feynman diagrams, the BFKL or QCD Pomeron may be viewed as a ladder diagram built by the exchange of two $t$-channel (Reggeized) gluons, see the left-hand side of Fig.~\ref{fig:cascade}. The sequence of parton spltting produces a gluon cascade which develops in ln$(1/x)$ space, and which is not strongly ordered in $k_t$, see the right-hand side of Fig.~\ref{fig:cascade}. There are phenomenological arguments (such as the small slope of the Pomeron trajectory\footnote{Recall that the slope $\alpha'_P \propto 1/\langle k_t^2 \rangle \propto R_{\rm Pom}^2$.}, the success of the Additive Quark Model relations, etc.) which indicate that the size of an individual Pomeron is relatively small as compared to the size of a proton or pion etc. Thus we may regard the cascade as a small-size `hot-spot' inside the colliding protons \cite{KMRhotspot}.

At LHC energies the interval of BFKL ln$(1/x)$ evolution is much larger than that for DGLAP ln$k_t^2$ evolution. Moreover, the data already give hints that we need contributions not ordered in $k_t$, $\grave{a}~ la$ BFKL, since typically DGLAP-based Monte Carlos, tuned to describe the Tevatron data, overestimate the observed $\langle k_t \rangle$ and underestimate the mean multiplicity observed at the LHC \cite{cms,atlas}. Further,  it is not enough to have only one Pomeron ladder exchanged; we need to include multi-Pomeron exchanges.

Basically, the picture is as follows. In the perturbative domain we have a single bare `hard' Pomeron exchanged with a trajectory $\alpha_P^{\rm bare}\simeq 1.3+\alpha'_{\rm bare}t$, where $\alpha'_{\rm bare} \lapproxeq 0.05$ GeV$^{-2}$. The transition to the soft region is accompanied by increasing absorptive multi-Pomeron effects as we go to smaller $k_t$ , such that a soft amplitude may be approximated by an {\it effective}  linear trajectory $\alpha^{\rm eff}_P \simeq 1.08+0.25t$ in the {\it limited} energy range up to Tevatron energies \cite{DL}.  This smooth transition of the Pomeron from hard to soft, that is from bare to effective,  is well illustrated, for example, by  
 the behaviour of the data for vector meson ($V=\rho, \omega, \phi, J/\psi$) production at HERA, $\gamma^*(Q^2)+p\to V(M)+p$. As $Q^2+M^2$ decreases from about 50 GeV$^2$ towards zero, the $s$ and $t$ dependence of the data reveal the trend that the effective intercept, $\alpha_P^{\rm eff}(0)$, decreases from about 1.3 to 1.1, and the slope, $\alpha_P^{\prime}$, increases from about zero to about 0.2 ${\rm GeV}^{-2}$.

\section{Partonic structure of the Pomeron --- the KMR model}

How can we implement the partonic model of the Pomeron in practice? This is achieved \cite{KMRnnn}\footnote{A two-channel eikonal is used, $i,k=1,2$.} from the partonic ladder structure of the Pomeron, $\Omega (y,\k,\b)$, generated by BFKL-like evolution in rapidity.
 $y$, with a simplified form of the kernel, $K$, but which incorporates the main features of BFKL: diffusion in ln$k_t^2$ and $\Delta=\alpha_P^{\rm bare}(0)-1\simeq 0.3$.   
At each step of the evolution in $y$, the value of the partonic $\k$ and the impact parameter, $\b$, can be changed.   

It is important to note that, in comparison with Reggeon Field Theory, $\Omega$ now depends on an extra variable, $\k$, that is the transverse momentum of the intermediate parton, in addition to the usual two `soft' Regge variables: $y={\rm ln}(1/x)$ and the impact parameter, $\b$, which is conjugated to the transverse momentum, $Q_t$, transferred through the entire ladder.

After including rescattering of the intermediate partons, we actually have 
two coupled evolution equations.
 One evolving up from the target $k$ at $y=0$, and one evolving down\footnote{The value of $Y_j$ accounts for the fact that at larger $k'_t$ a smaller rapidity interval is available for the evolution.} from the beam $i$ at $y'=Y_j-y=0$ with $Y_j={\rm ln}(s/k^{'2}_t)$,
\be
\frac{\partial \Omega_k(y)}{\partial y}~=~\int \frac{d^2\kk}{\pi k^{'2}_t}~{\rm exp}(-\lambda[\Omega_k(y)+\Omega_i(y')]/2)~K(\k,\kk)~\Omega_k(y).
\label{eq:e15}
\ee
\be
\frac{\partial \Omega_i(y')}{\partial y'}~=~\int \frac{d^2\kk}{\pi k^{'2}_t}~{\rm exp}(-\lambda[\Omega_i(y')+\Omega_k(y)]/2)~K(\k,\kk)~\Omega_i(y'),
\label{eq:e16}
\ee
where, for clarity, we have suppressed \footnote{In general, the kernel $K$ depends on the difference $\b-\bb$. However, the $\b$ dependence is proportional to the slope $\alpha'$ of the bare Pomeron trajectory. Asymptotically the BFKL approach predicts $\alpha' \to 0$; and indeed, analyses of soft data find it to be very small. Thus, for simplicity, we have not shown this variation in eqs. (\ref{eq:e15}, \ref{eq:e16}). Then the only $\b$ dependence of $\Omega$ comes from the starting distribution of the evolution, and not from the $\b$ dependence of $K$.} the $\k$ labels of the $\Omega$'s. The inclusion of the $k_t$ dependence is crucial for the transition from the hard to the soft domain. The absorptive (exponential) factors\footnote{Since we are dealing with the {\it amplitude} $\Omega$, and not with the cross section, we use here $\exp(-\lambda\Omega/2)$ and not $\exp(-\lambda\Omega)$. The parameter $\lambda$ determines the strength of the triple- (and multi-) Pomeron couplings in terms of the Pomeron-proton coupling. In practice, a slightly different form of the absorptive factor \cite{KMRnnn} is preferred by the data, which is consistent with the AGK cutting rules \cite{AGK}. } in the equations embody the result that there is less screening for larger $k_t$. The coupled evolution equations may be solved iteratively to give $\Omega(y,\k,\b)$, for a whole range of fixed values of $\b$.

Recall that the absorptive factors, exp$(-\lambda\Omega/2)$, in (\ref{eq:e15}, \ref{eq:e16}) lead to a strong suppression of the low $k_t$ domain, which introduces an effective, dynamically generated, infrared cutoff $k_t>k_{\rm min}$, whose value increases with energy. Therefore the final high energy results are not too sensitive to the value of the original cutoff that is included artificially in the propagators ($1/k^2 \to 1/(k^2+Q_0^2)$) in order to protect the infrared singularity.

In principle, knowledge of $\Omega_k(y,\k,\b)$ allows a good description of all soft and semi-hard high-energy $pp$ data, such as $\sigma_{\rm tot},~d\sigma_{\rm el}/dt,~d\sigma_{\rm SD}/dtdM^2,..$; and reliable estimates of the survival factors of large rapidity gaps \cite{KMRsurvival}; and even semi-quantitative estimates of PDFs and diffractive PDFs at low $x$ and low scales. Moreover, such a model \cite{KMRnnn} has only a few physically motivated parameters. Specifically, the parameters are the bare Pomeron intercept $\Delta\equiv \alpha_P(0)-1$ (expected to be about 0.35) and slope $\alpha_P^{\prime}$ (expected to be small); $d$ which controls the diffusion in ln$k_t^2$; $\lambda$; the initial gluon density, $N$; and the parameters which specify the diffractive eigenstates.

Examples of topical predictions of the model are that the rapidity plateau increases with collider energy as $d\sigma/dy \sim s^{0.2}$, just like the LHC data from 0.9 to 7 TeV; or that the gap survival for double diffractive (exclusive) SM Higgs production, $pp \to p+h+p$ at 14 TeV the LHC\footnote{If the outgoing protons in this process are measured far from the interaction point then the mass of the SM Higgs particle can be determined, via missing mass, to an accuracy of about 1 GeV.  If $M_h$ is less than about 140 GeV, this process offers a chance to study the $h \to b\bar{b}$ decay, since the QCD $b\bar{b}$ background is heavily suppressed by a $J_z=0$ selection rule \cite{J0}.  Such exclusive events are very clean, but $\sigma\sim 2$ fb \cite{KMRpros}, and events are contaminated by `pile-up'. With moderate pile-up, very accurate event timing can select the exclusive events. An advantage of such a process is that, in the case of MSSM, the $h$ and $H$ Higgs of may be both observable, and the cross sections may be enhanced. } is about 0.02. Exclusive (low multiplicity) processes, like
 $d\sigma_{\rm el}/dt$ and $pp \to p+h+p$, may be calculated analytically, whereas in inclusive cases we need a Monte Carlo to account for hadronization effects. In particular, we need to allow for hadronization to describe the ATLAS measurements,  Fig.~{\ref{fig:Newman}, of $d\sigma/d(\Delta\eta)$ for $\Delta\eta<5$; although  for larger gaps the triple-Pomeron diagram with absorption corrections will suffice, see the discussion in \cite{KKMRZ}. 
\begin{figure}[htb]
\begin{center}
\vspace{-1.7cm}
\resizebox{0.8\textwidth}{!}{
\includegraphics{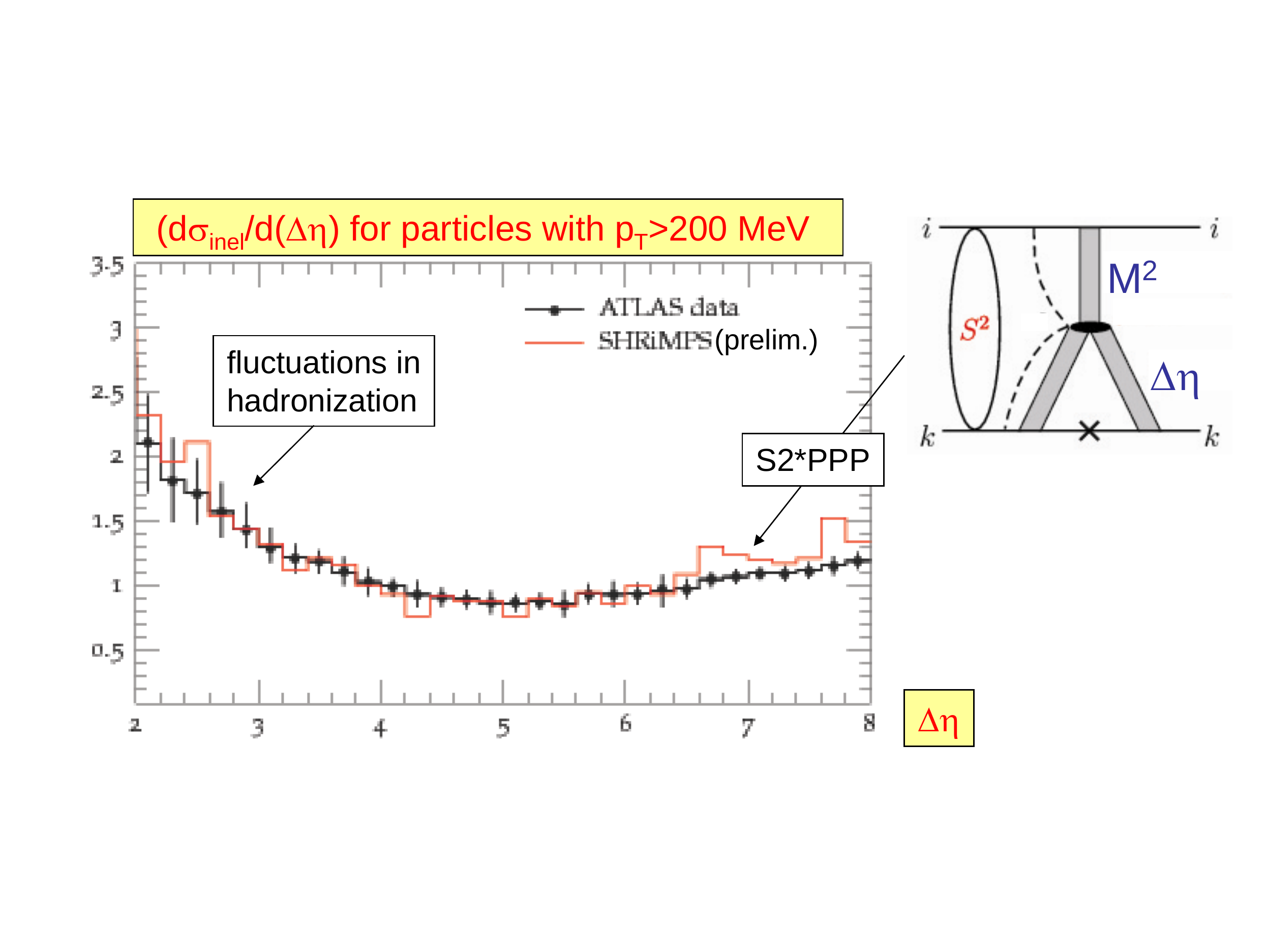}\hspace{1cm}
}
\vspace{-2cm}
\caption{\sf The inelastic cross section differential in rapidity gap size $\Delta\eta$ for particles with $p_T>200$ MeV. The data are from \cite{atlas1}. The triple-Pomeron diagram, with multi-Pomeron corrections, can describe the high $\Delta\eta$ data \cite{KMRopacity}, but hadronization effects are evident at smaller $\Delta\eta$.}
\label{fig:Newman}
\end{center}
\end{figure}

\section{The SHRiMPS Monte Carlo}

Most of the existing general purpose Monte Carlo event generators (Pythia, Herwig, Herwig++, Sherpa), \cite{MC}, use collinear factorisation and standard $2\to 2$ QCD matrix elements supplemented with a factorised spatial dependence as the basis for their eikonals, $\Omega$.  In order to avoid potential problems in the infrared region, however, these matrix elements are {\it either} cut at some low value of minimal $k_t$ of the outgoing partons, {\it or} the soft behaviour of the $t$-channel and $\alpha_S$ are modified ($|t| \to |t|+q_0^2$) to render the result finite.  In the former case, when the `hard' eikonal is cut at $k_0$, it is further supplemented with a soft one, $\Omega=\Omega_{\rm hard}+\Omega_{\rm soft}$, to account for the contribution from the (infrared) region $k_t<k_0$.  In any case, the simulation typically then uses total cross sections etc. as input parameters and concentrates on producing the correct particle distributions.

Here, we seek a Monte Carlo that describes all aspects of minimium bias events --- total, differential elastic cross sections, proton dissociation, diffraction, jet production etc. --- in a {\it unified} framework, capable of modelling exclusive final states.
In order to obtain such a Monte Carlo, which includes a description of soft physics and diffraction as well as jet physics in a self-consistent way, the KMR model has been incorporated into the SHERPA framework \cite{SHRiMPS}. Recall that the KMR model is based on the bare QCD Pomeron, with absorptive multi-Pomeron rescattering corrections, that increase as we continue from the {\it hard} to the {\it soft} domain.  For this reason we call the Monte Carlo `SHRiMPS', standing for {\bf S}oft {\bf H}ard {\bf R}eactions {\bf i}nvolving {\bf M}ulti {\bf P}omeron {\bf S}cattering.

Let us outline the construction of the Monte Carlo. First, we solve the coupled evolution equations in rapidity $y$ to generate $\Omega_{ik}(y,k_t,b)$, having specified the boundary conditions of the diffractive (Good-Walker) eigenstates. These eigenstates specify the elastic and quasi-elastic scattering.  For a detailed simulation of the inelastic state we must select the number of ladders to be exchanged. We take the number to be distributed according to a Poisson distribution, with parameter $\Omega_{ik}(b)$.  Having decided on the number of (primary) ladders to be exchanged, the incoming protons are dissolved into a valence quark, a (non-interacting)valence diquark and gluons. A random pair of partons from these beam dissociations is chosen to exchange the next ladder. Gluon emmisions are generated in the ladder according to a Markov chain, ordered in rapidity, with a pseudo-Sudakov form factor. The $t$-channel propagators in the ladder are reggeized gluons that come in different colour states; only singlets and octets are considered. For each $t$-channel propagator we select the probability of a colour singlet,
\begin{equation}
{\cal P}_1~=~(1-e^{-\delta\Omega/2})^2,
\end{equation}
where $\delta\Omega$ is known from the solution of (\ref{eq:e15}, \ref{eq:e16}).
We take ${\cal P}_8 = 1-{\cal P}_1$. Each gluon emission leads to two new propagators: allowed combinations are ${\cal P}_1{\cal P}_8, {\cal P}_8{\cal P}_1$ and ${\cal P}_8{\cal P}_8$.
The formation of singlet propagators gives rise to rapidity gaps being associated with `elastic rescattering'. Simultaneously the rescattering may lead to the inelastic interaction of secondaries, producing new ladders with Poisson probabilities $e^{-\delta\Omega}(\delta\Omega)^n/n!$. The single gluon emission process is iterated until the active interval becomes a colour singlet or no further emissions are kinematically allowed in the rapidity interval.  Finally we implement the `usual' parton shower, plus hadronization, plus hadron decays, plus QED to produce the final scatter of observed particles.

\begin{figure}[htb]
\begin{center}
\vspace{-0.5cm}
\resizebox{0.8\textwidth}{!}{
\includegraphics{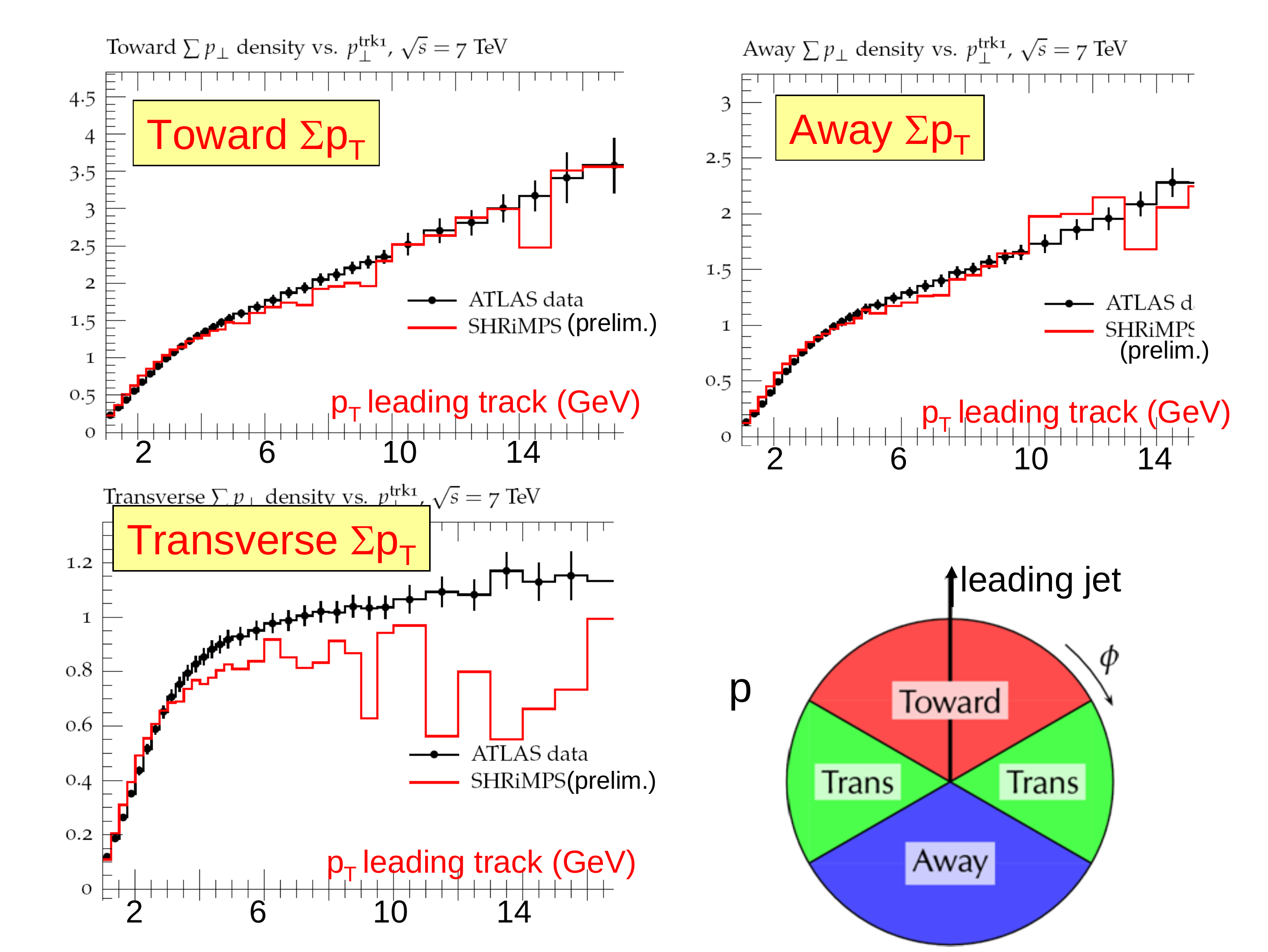}\hspace{1cm}
}
\caption{\sf The values of $\sum p_T$ in the toward, away and transverse regions to the direction of a leading jet of given $p_T$(jet). The data are from \cite{atlas2}.}
\label{fig:FTA}
\end{center}
\end{figure}

The Monte Carlo, which has relatively few parameters (several of which are physically motivated), is tuned to an extensive range of `soft' high energy data. Two examples of the preliminary description of data are shown in Figs.~\ref{fig:Newman} and \ref{fig:FTA}.


\begin{thebibliography}{99}

\bibitem{review} See, for example, A.B. Kaidalov, Phys. Rept. {\bf 50}, 157 (1979); \\ 
A.D. Martin, M.G. Ryskin and V.A. Khoze, Acta Phys. Polon. {\bf B40}, 1841 (2009).

\bibitem{RFT} V.N. Gribov, Sov. Phys. JETP {\bf 26}, 414 (1968)

\bibitem{book} for a recent detailed review see 
V.S.~Fadin, B.L.~Ioffe and L.N.~Lipatov,
{\it in} Quantum Chromodynamics
(Camb. Univ. Press, 2010).

\bibitem{bfklresum} M. Ciafaloni, D. Colferai and G. Salam, Phys. Rev. {\bf D60}, 114036 (1999);\\
V.A.~Khoze, A.D.~Martin, M.G.~Ryskin and W.J. Stirling, Phys. Rev. {\bf
D70}, 074013 (2004).

\bibitem{KMRhotspot}  M.G. Ryskin, A.D. Martin and V.A. Khoze, J. Phys. G {\bf 38}, 085006 (2011).

\bibitem{cms} CMS Collaboration,
  Phys.\ Rev.\ Lett.\  {\bf 105}, 022002 (2010).
  
\bibitem{atlas}  ATLAS Collaboration, Phys. Rev. {\bf D83}, 112001 (2011). 
 
\bibitem{DL} A. Donnachie and P.V. Landshoff, Phys. Lett. {\bf B296}, 227 (1992).

\bibitem{KMRnnn} M.G. Ryskin, A.D. Martin and V.A. Khoze, Eur. Phys. J. {\bf C71}, 1617 (2011).

\bibitem{AGK} V.A. Abramovsky, V.N. Gribov and O.V. Kancheli, Sov. J. Nucl. Phys. {\bf 18},308 (1974).
\bibitem{KMRsurvival}  M.G. Ryskin, A.D. Martin and V.A. Khoze, Eur. Phys. J. {\bf C60}, 265 (2009).

\bibitem{J0} 
V.A. Khoze, A.D. Martin, and M.G. Ryskin,
 hep-ph/0006005, Proc. 8th DIS 2000 Workshop;\\
  V.A.~Khoze, A.D.~Martin and M.G.~Ryskin,
 Eur.\ Phys.\ J. {\bf C19}, 477 (2001)
  [Err: {\bf C20}, 599 (2001)].

\bibitem{KMRpros} V.A. Khoze, A.D. Martin and M.G. Ryskin, Eur. Phys. J. {\bf C23}, 311 (2002).

\bibitem{atlas1} ATLAS Collaboration, Eur. Phys. J. {\bf C72}, 1926 (2012).

\bibitem{KMRopacity}  M.G. Ryskin, A.D. Martin and V.A. Khoze, Eur. Phys. J. {\bf C72}, 1937 (2012).

\bibitem{KKMRZ} V.A. Khoze et al., Eur. Phys. J. {\bf C69}, 85 (2010).

\bibitem{MC} See the review in  
  A.~Buckley {\it et al.},
  Phys.\ Rept.\  {\bf 504}, 145 (2011).

\bibitem{SHRiMPS} SHRiMPS Monte Carlo, H. Hoeth, V.A. Khoze, F. Krauss {\it et al}., in preparation.

\bibitem{atlas2} ATLAS Collaboration,  
  Phys.\ Rev.\ D {\bf 83}, 112001 (2011).


\end{thebibliography}
\end{document}